# Resting-state functional connectivity in late-life depression: higher global connectivity and more long distance connections

Iwo Jerzy Bohr[1], Eva Kenny[2,4,], Andrew Blamire[3,4], John T O'Brien[2], Alan J Thomas[2], Jonathan Richardson[2], Marcus Kaiser*[1,5,6]

1 School of Computing Science, Claremont Tower, Newcastle University, Newcastle upon Tyne, UK
2 Institute for Ageing and Health, Campus for Ageing and Vitality, Newcastle University, Newcastle upon Tyne, UK
3 Institute of Cellular Medicine, Newcastle University, Newcastle upon Tyne, UK
4 Newcastle Magnetic Resonance Centre, Campus for Ageing and Vitality, Newcastle University, Newcastle upon Tyne, UK
5 Institute of Neuroscience, Newcastle University, Newcastle upon Tyne, UK
6 Department of Brain and Cognitive Sciences, Seoul National University, Seoul Korea

Corresponding author:
Marcus Kaiser
Newcastle University
School of Computing Science
Claremont Tower
Newcastle upon Tyne NE1 7RU
UK
Phone: +44 191 222 8161
FAX: +44 191 222 8232

marcus.kaiser@ncl.ac.uk

**Abstract**
Functional magnetic resonance imaging (fMRI) recordings in the resting-state (RS) from the human brain are characterized by spontaneous low-frequency fluctuations (SLFs) in the blood oxygenation level dependent (BOLD) signal that reveal functional connectivity (FC) via their spatial synchronicity. This RS study applied network analysis to compare FC between late-life depression (LLD) patients and control subjects. Raw cross-correlation matrices (CM) for LLD were characterized by higher functional connectivity. We analysed the small-world and modular organization of these networks consisting of 110 nodes each as well as the connectivity patterns of individual nodes of the basal ganglia. Topological network measures showed no significant differences between groups. The composition of top hubs was similar between LLD and control subjects, however in the LLD group posterior medial parietal regions were more highly connected compared to controls. In LLD, a number of brain regions showed connections with more distant neighbours leading to an increase of the average Euclidean distance between connected regions compared to controls. In addition, right caudate nucleus connectivity was more diffuse in LLD. In summary, LLD was associated with overall functional connectivity strength and changes in the average distance between connected nodes, but did not lead to global changes in small-world or modular organization.

# 1. Introduction

Late-life depression (LLD) is a common psychiatric disorder that typically occurs after 60 years of age. Prevalence rates can range from 1-4% for major and up to 13% for minor depression. Whereas volume reductions in cortical- and subcortical regions can be found, it is unclear what the consequences for cognitive functions may be. In this study, resting-state (RS) functional magnetic resonance imaging (rs-fMRI) is used to observe functional connectivity (FC) indicating correlated activity patterns in different parts of the brain (Fox and Raichle, 2007;Auer, 2008). In rs-fMRI, spontaneous low-frequency fluctuations (SLFs, 0.01-0.1 Hz) occur in the BOLD signal in globally distributed brain areas, which form functionally related networks, termed RS networks (RSNs) (Fox and Raichle, 2007;Auer, 2008;van den Heuvel and Hulshoff Pol, 2010). Default Mode Network (DMN) SLFs are negatively correlated with tasks requiring focused attention (Raichle et al., 2001;Greicius et al., 2003;Buckner et al., 2008). The DMN includes the ventral medial prefrontal cortex and the posterior cingulate cortex also stretching to the precuneus and intraparietal lobule. Primary sensory or motor regions are absent from the DMN (Buckner et al., 2008).

There are two main approaches to investigate FC: hypothesis-driven and data-driven. Hypothesis-driven approaches involve the selection of a seed and functional connectivity is investigated with either a pre-defined brain region(s) or all other brain voxels by correlation of the SLF in the seed region with the other brain regions. In contrast, data-driven approaches are not based on any *a priori* hypothesis about the importance of specific brain areas and look into patterns emerging as a result of the analysis of the activity in the brain as a whole. Compared to a previous hypothesis-driven LLD study (Kenny et al., 2010), we here use a data-driven approach.

We apply network analysis to characterize whole-brain changes in functional connectivity. Network analysis provides a range of tools for studying brain regions (treated as nodes of the network) and interactions (edges) (Sporns et al., 2004;Stam and Reijneveld, 2007;Reijneveld et al., 2007 ;Bullmore and Bassett, 2010;Kaiser, 2011).

Brain connectivity was found to show properties of Small-World (SW) networks (Watts and Strogatz, 1998) for various techniques (fMRI, EEG, tract tracing) and various species and levels of organisation (*C. elegans*, rat, cat, macaque, human). SW networks are characterized by a relatively small number of links that must be passed to "travel" between a pair of nodes. This may be expressed as the characteristic path length; *L*. Small-world networks also display high values of interconnectedness of neighbouring nodes (high clustering coefficient, *C*). Therefore, SW properties in brain networks ensure efficient processing while reducing the total cost of wiring (Bassett and Bullmore, 2006;Kaiser and Hilgetag, 2006).

Brain networks also show a hierarchical modular organization (Bassett et al., 2010) and contain highly-connected nodes or hubs (Hagmann et al., 2008). Hubs are often critical for the structural and functional integrity of a network. In many cases they play a role of "bridges" between nodes and often between clusters, thus assuring a low value of characteristic path length. For resting-state functional connectivity, most hubs are part of the DMN; for example, the precuneus or parietal and medial prefrontal cortex (Buckner et al., 2009). Hubs have been characterized by a high number of long distance connections (Achard et al., 2006) and a tendency towards an inverse relationship between Euclidean distances and fluctuation frequency (Salvador et al., 2005). A number of diseases have an impact on functional connectivity (Buckner et al., 2008;Bassett and Bullmore, 2009), including Alzheimer's disease (Supekar et al., 2008), schizophrenia (Liu et al., 2008) and depression (Zhang et al., 2011). It has been proposed that neurodegenerative diseases specifically target critical network components, such as hubs and sets of hubs (Buckner et al., 2009;Seeley et al., 2009); therefore, alterations in RSNs might be causes rather than consequences of these disorders.

Depression can be categorized as either major or minor based on duration, number of symptoms and severity. Five of the core symptoms must be present for at least two weeks for a diagnosis of major depression to be fulfilled; one symptom must be depressed mood or loss of interest/enjoyment in everyday activities (anhedonia). The symptoms must have a significant impact on occupational and/or social functioning in order for criteria to be fulfilled (Meunier et al., 2009). LLD, typically occurring after 60 years of age, can cause great suffering in the elderly and reduce their quality of life. LLD is frequently comorbid with physical illnesses, for example it is common in patients recovering from myocardial infarction (American Psychiatric Association, 1994), and when present can delay recovery and lengthen hospital stay. Compared to other diseases, there are few studies on the relationship between FC and LLD. Findings have varied with some reporting increased connectivity (Kenny et al., 2010), others increased and decreased connectivity (Yuan et al., 2008), and others decreased only (Aizenstein et al., 2009).

In this study, we measured resting-state functional connectivity using a data-driven analysis approach thus extending the findings from a previous hypothesis-driven study that used a seed correlation analysis approach (Kenny et al., 2010). Note that we selected a group of patients not displaying symptoms of depression at the time of the investigation since our aim was to look at the traits rather than the state of the disease (see also Methods and Discussion for more details on this matter).

## 2. Methods

### 2.1. Participants

This study involved 30 subjects: 14 with a history of major depression (LLD group) and 16 (age-matched) control individuals. Patients were recruited from consecutive referrals to Newcastle and Gateshead Old Age Psychiatry Services. All subjects were aged 65 years or older. Control participants were recruited by advertisement; none of the control subjects had past or present history of depression. A full neuropsychiatric assessment was conducted including family history of depression, previous psychiatric history, medical history and current medication. Current depression severity was rated using the Montgomery-Åsberg Depression rating scale (MADRS) (Montgomery and Asberg, 1979). Depressed subjects were required to fulfil DSM-IV criteria for a life-time diagnosis of major depressive episodes (American Psychiatric Association, 1994). Patients were assessed by senior psychiatrists in the NHS and then by a senior research psychiatrist (JR) who applied DSM criteria. All psychiatrists were MRCPsych and fully trained, equivalent of Board Certified in US. Comorbidity was assessed by physical examination, including cardiovascular and ECG, by JR.

All subjects were also assessed on the Mini-Mental State Examination (MMSE) to exclude the presence of dementia (Folstein et al., 1975). For all participants, the following exclusion criteria applied: dementia or MMSE<24 (absence of dementia in referred subjects was confirmed by AV), current use of a tricyclic antidepressant, comorbid or previous drug or alcohol misuse, previous head injury, previous history of epilepsy, previous transient ischaemic attack (TIA) or stroke, a carotid bruit on physical examination, myocardial infarction (MI) in the previous 3 months, a depressive episode in the previous three months, or contraindication to MRI screening. The study was approved by the Newcastle and North Tyneside Research Ethics Committee and all subjects gave verbal and written consent.

Table 1 shows the clinical characteristics of the study subjects. Groups were comparable for gender ($\chi^2$ = 1.2, df = 1), age, and MMSE score. Mean MADRS score for LLD subjects was 7.5, indicating that most had recovered from their episode of depression by the time of scanning. Mean age at onset of depression was 49.8 years and the number of previous episodes of depression was 2.6. At the time of the study, four LLD subjects were taking antidepressants (citalopram and lofepramine), two were taking antipsychotics (flupenthixol and prochlorperazine), one was taking non-benzodiazepine hypnotic (zopiclone), and one an antiepileptic drug (carbamazepine).

**Table 1** Demographic and neuropsychological data of controls and late-life depression (LLD) patients.

| Demographic/ Neuropsychological Data | Controls | LLD | $p$ value |
|---|---|---|---|
| N | 16 | 14 | |
| Sex (M:F) | 10:6 | 8:6 | 0.27 [a] |
| Age (Yrs) | 75.8±7.8 | 76.6±7.7 | 0.77 [b] |
| MMSE | 28.9±1.2 | 28.0±1.9 | 0.27 [b] |
| MADRS | | 7.5±4.7 | |
| Age at Onset of Depression | | 49.8±18.8 | |
| No. of Previous Episodes of Depression | | 2.6±2.1 | |

Values expressed as mean ± standard deviation. MADRS = Montgomery-Asberg Depression Rating Scale; MMSE = Mini Mental State Examination.
[a] The p value was calculated using Chi-Square Test.
[b] The p values were calculated using the Independent-Samples t-test.

## 2.2. Image acquisition and pre-processing

Images were acquired using a 3 Tesla scanner (Intera Achieva, Philips Medical Systems, The Netherlands), with an 8-channel head coil. Conventional T1-weighted 3-dimensional scans: magnetisation-prepared rapid acquisition with gradient echo (MPRAGE) were collected for anatomical mapping. Sagittal slices were acquired of thickness = 1.2 mm, voxel size = 1.15 x 1.15 mm, repetition time (TR) = 9.6 ms, echo time (TE)= 4.6 ms, flip angle = 8˚, SENSE factor = 2.
Subjects were instructed to lie still in the scanner, to keep their eyes closed but not to fall asleep while resting-state images were collected using a gradient-echo echo-planar imaging (GE-EPI) sequence with the following parameters: TE =40 ms, TR=3000 ms, flip angle 90˚, 25 contiguous axial slices of 6 mm thickness, field of view (FOV)=260×260 mm, in-plane resolution 2×2 mm. A total of 128 volumes were collected per subject, with a total scan time of 6.4 minutes. As previously shown, this number of volumes is sufficient to obtain stable network features (van Wijk et al., 2010) .

Images were pre-processed using FSL (Smith et al., 2004;Woolrich et al., 2009) to correct for subject motion (MCFLIRT) (Jenkinson et al., 2002)  and to extract the brain from non-neural tissue (BET) (Smith, 2002).  We also applied spatial smoothing (5 mm full width at half maximum) and high-pass temporal filtering (cut-off = 125s; FEAT, version 5.92). To account for age-related anatomical changes, such as ventricular enlargement or gyri shrinking, anatomical scans were transformed to standard space and averaged to create a subject-specific template for registering our functional imaging data.

Anatomical T1 images were segmented into grey matter, white matter and cerebrospinal fluid (CSF) using SPM5 (Wellcome Department of Imaging Neuroscience Group, London, UK) implemented in Matlab R2009a (Mathworks Inc., Natick, MA, USA), and total intracranial volume was calculated from the sum of the three components. We did not find a significant difference in brain volumes between controls and the LLD group using unpaired two-sample t-test.

## 2.3. Functional connectivity analysis workflow

Parcellation was performed using FSL and was based on the Harvard-Oxford Probabilistic MRI Atlas (HOA). This involved extracting 48 cortical and seven subcortical regions (thalamus, caudate, putamen, pallidum, amygdala, nucleus accumbens and hippocampus) from the respective parts of the atlas, thus totaling  in 110 brain regions in two hemispheres. Note, that network properties relate to the number of nodes in a network (Echtermeyer et al., 2011) and we therefore chose 110 nodes to be comparable with majority of previous whole-brain networks studies based on macroanatomical atlases; see for instance a recent paper indicating similar results of FC analysis using three types of macroanatomical atlases (Spoormaker et al., 2012).  FLIRT was used to register structural images to functional images, averaging over each ROI for each volume and demeaned time series for each area extracted. Using custom scripts in Matlab (Release 2009a,), data from each individual were placed in one temporary matrix for each subject (n × m; n=number of nodes=110, m=number of scans=128), global signal removed (mean BOLD signal subtraction for all nodes) and transformed into connectivity matrix (CM) representing all 110 nodes. Self-correlations, across the diagonal of CM, were disregarded.

## 2.4. Network analysis

The raw CM represents weighted un-directed graphs. We observed the average correlation between all pairs of nodes (cross-correlation matrix). This procedure was applied to (a) the raw CMs, (b) CMs with negative correlation values set to zero, and (c) CMs with a percentage of top positive correlations remaining and all other correlations set to zero. The latter CMs were used to generate binary networks, setting all non-zero values to 1. For this, the 20% of top correlations (Pearson $r$-values) were considered as functionally connected nodes. Such thresholding led to equal edge densities in all subjects, which is required for comparisons of network topology. Using different edges densities, e.g. by using a constant correlation value as threshold for all subjects, would otherwise directly influence network features (Jenkinson and Smith, 2001). In addition, we chose a 20% edge density to be in line with what would be expected from the edge density of the underlying structural connectivity which ranges from 10-30% (van Wijk et al., 2010). The 20% edge density led to an average correlation threshold of $r$=0.28 which is close to the threshold in an earlier study (Kaiser, 2011).

We calculated several topological features for the thresholded binary networks (see supplementary information or (Achard et al., 2006;Bassett and Bullmore, 2006) for more details): First, the characteristic path length *L*, which is the average number of connections that have to be crossed to go from one node to another on the shortest-possible path. Second, the clustering coefficient *C*, that defines what proportion of neighbours (nodes which are directly connected to a node) are connected to each other. Small-world networks are characterized by a clustering coefficient that is much higher than for a randomly connected network while the characteristic path length is still comparable to that of a random network (Kaiser, 2011). A way to assess the extent of such a small-world organization is the small-worldness σ as defined by σ = $C L_r/(C_r L)$ where $L_r$ and $C_r$ are the characteristic path length and clustering coefficient of a random benchmark network, respectively (Watts and Strogatz, 1998). Third, we observed the modularity *Q* that determines the degree to which a network is organized into distinct modules. In addition to topological changes, we also searched for changes in spatial organization. The three-dimensional location of a node was given by the centre of mass of a region's coordinates in FSL. The Euclidean distance between connected nodes was used as an approximation of the connection reach.

  *-- Fig 1 about here –*

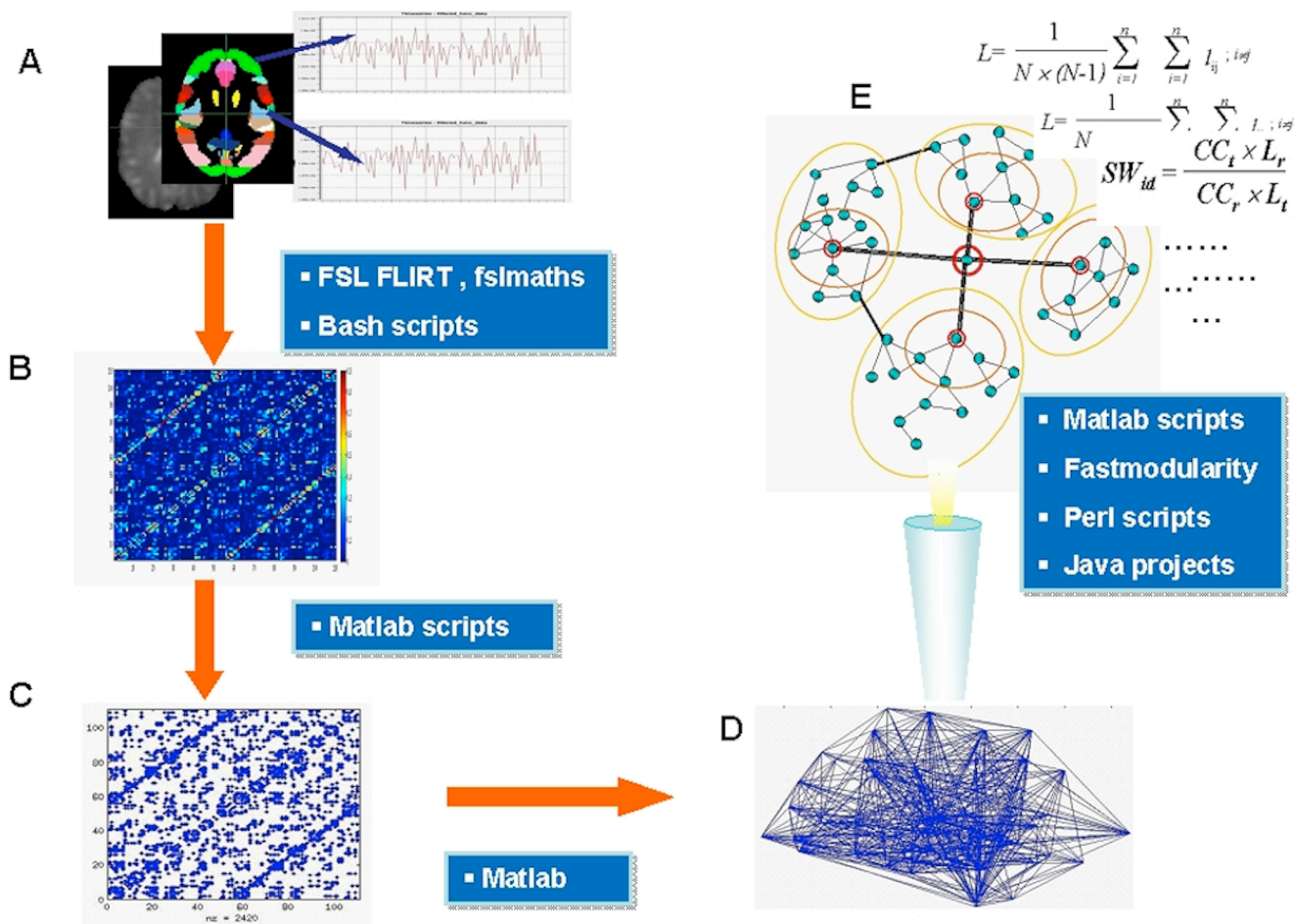

## 2.5. Statistical analysis

Values for metrics of global FC are quoted as mean ± SD (standard deviation). Two-sample t-tests were performed to check for statistical differences of single measures between the two groups, with p<0.05 thresholds for significance of global measures and p<0.01 for node-wise analysis (corrected for multiple comparisons; number of nodes: 110). All correlations were tested with Pearson coefficient (r) and with t-test (n-2 degree of freedom; n=number of rows in a correlation matrix) for significance. To correct for multiple comparisons in the case of node-wise analysis, we used non-parametric permutation tests (Humphries and Gurney, 2008), (5,000 iterations) with a False Discovery Rate (FDR) of 5% (implemented by Dr Cheol Han in a Matlab script). Analysis was performed using SPSS (version 15.0.1) and Matlab.

## 3. Results
### 3.1. Global network

LLD showed a higher association at a global level as measured by the cross-correlations $r$, averaged across all subjects in each group (p=0.037): $r_{av}$=0.006483±0.010662 vs. $r_{av}$=0.000411±0.00482 for patients and controls, respectively. There was no difference between groups after setting negative correlations to zero.

Global network measures for binary networks ($L_{av}$, $C_{av}$, $\gamma$, $\lambda$, $\sigma$ and $Q$) yielded very similar values for both groups (Table 2). The values for average characteristic path length $L_{av}$ were similar for controls and LLD participants (2.20±0.14 and 2.20±0.19, respectively), as was the value for average clustering coefficient $C_{av}$ (0.58±0.05 in both groups). The values of $L$ and $C$ suggest a small-world architecture of the FC networks. This is confirmed by high values of small-worldness $\sigma$ (2.30±0.07 and 2.27±0.13, respectively) and consistent with the ratio of path lengths $\gamma$ (2.78±0.25 and 2.80±0.21) and of clustering coefficients $\lambda$ (1.20±0.08 and 1.23±0.11) between FC and benchmark random networks with the same number of nodes and edges. These findings indicate that the LLD group preserved small-world and modular characteristics despite the mental changes caused by depression. Interestingly, there was no correlation between clustering coefficient $C$ and modularity $Q$ in LLD whereas these two measures of modular organization strongly correlate with each other within controls ($r$=0.6; p<0.05).

**Table 2** Summary of global aggregate measures in the two groups (means ± SD)

|  | Controls | LLD |
|---|---|---|
| Grand mean for row cross correlation matrices | 0.000411±0.0048 | 0.00648±0.011* |
| Grand mean for thresholded cross correlation matrices | 0.454229±0.05358 | 0.465225±0.071135 |
| Characteristic path length (L) | 2.20 ±0.14 | 2.20± 0.19 |
| Clustering coefficient (CC) | 0.58± 0.05 | 0.58± 0.05 |
| γ | 2.78±0.25 | 2.80±0.21 |
| λ | 1.20±0.08 | 1.23±0.11 |
| Small-world Index (σ) | 2.30±0.07 | 2.27±0.13 |
| Modularity (Q) | 0.39 ±0.04 | 0.37± 0.03 |

* Significantly higher (p<0.05, t-test)

## 3.2. Local regions

Topological measures, when applied to each node separately, did not yield significant inter-group differences ($L_i$, $C_i$, $\sigma$ and k). In contrast, mean Euclidean distances (ED) of neighbours changed for several nodes (Figure 2). The average distance between connected nodes in LLD patients was significantly higher than in the control group for 14 regions and significantly lower than in the control group for two brain areas in the left hemisphere: Middle Temporal Gyrus, and Supramarginal Gyrus (all significant differences at 5% FDR).

Composition of the top 15 hubs (Table 3) did not yield significant differences. In addition there was a great deal of inter-individual variance in the two groups as far as composition of this core is concerned. The hub that occurred most consistently (60% in both groups) within the core was the posterior supra-marginal gyrus (PSG). In controls the second most frequently occurring hub was a frontal area: middle frontal gyrus (MFG), whereas in LLD it was an anterior division of the SG (53%) coming in at third position in frequency ranking (33%), slightly ahead of the posterior cingulate cortex PCC (27%;), which was less frequent in controls (20%). Therefore a tendency towards more medial-parietal areas as the most frequent hubs in LLD was observed. This was in contrast to controls in which frontal and temporal areas seemed to dominate.

**Table 3** List of 15 top hubs for controls and LLD group. L: right, R: right, k: degree centrality values, C: cortex, G: gyrus, ant.: anterior, pariet.: parietal.

| Controls | | LLD | |
| --- | --- | --- | --- |
| Area name | k | Area name | k |
| R- Paracingulate G | 38 | R-Paracingulate G | 46 |
| R- Supramarginal G, posterior division | 37 | L-Supramarginal G, ant. division | 43 |
| L- Middle Frontal G | 36 | L-Paracingulate G | 42 |
| L-Paracingulate G | 36 | L-Central Opercular C | 42 |
| R-Lateral Occipital C, superior division | 36 | L-Precentral G | 41 |
| L-Cingulate G, ant. Division | 34 | R-pariet. Operculum | 41 |
| L-Central Opercular C | 34 | L- Supramarginal G, posterior division | 39 |
| R-Angular G | 33 | L-Cingulate G, ant. Division | 39 |
| L-pariet. Operculum | 33 | L-pariet. Operculum | 39 |
| L-Supramarginal G, posterior division | 32 | R-Lateral Occipital C, superior division | 39 |
| L-Putamen | 32 | R-Supramarginal G, posterior division | 38 |
| R_Frontal Pole | 32 | R- Angular G | 38 |
| R- Juxtapositional Lobule C (formerly Supplementary Motor C) | 32 | R-Cingulate G, ant. division | 38 |
| R-Cingulate G, ant. Division | 32 | R-Precuneus C | 37 |
| L-Insular C | 31 | L-Lateral Occipital C, superior division | 36 |

*-- Fig 2 about here –*

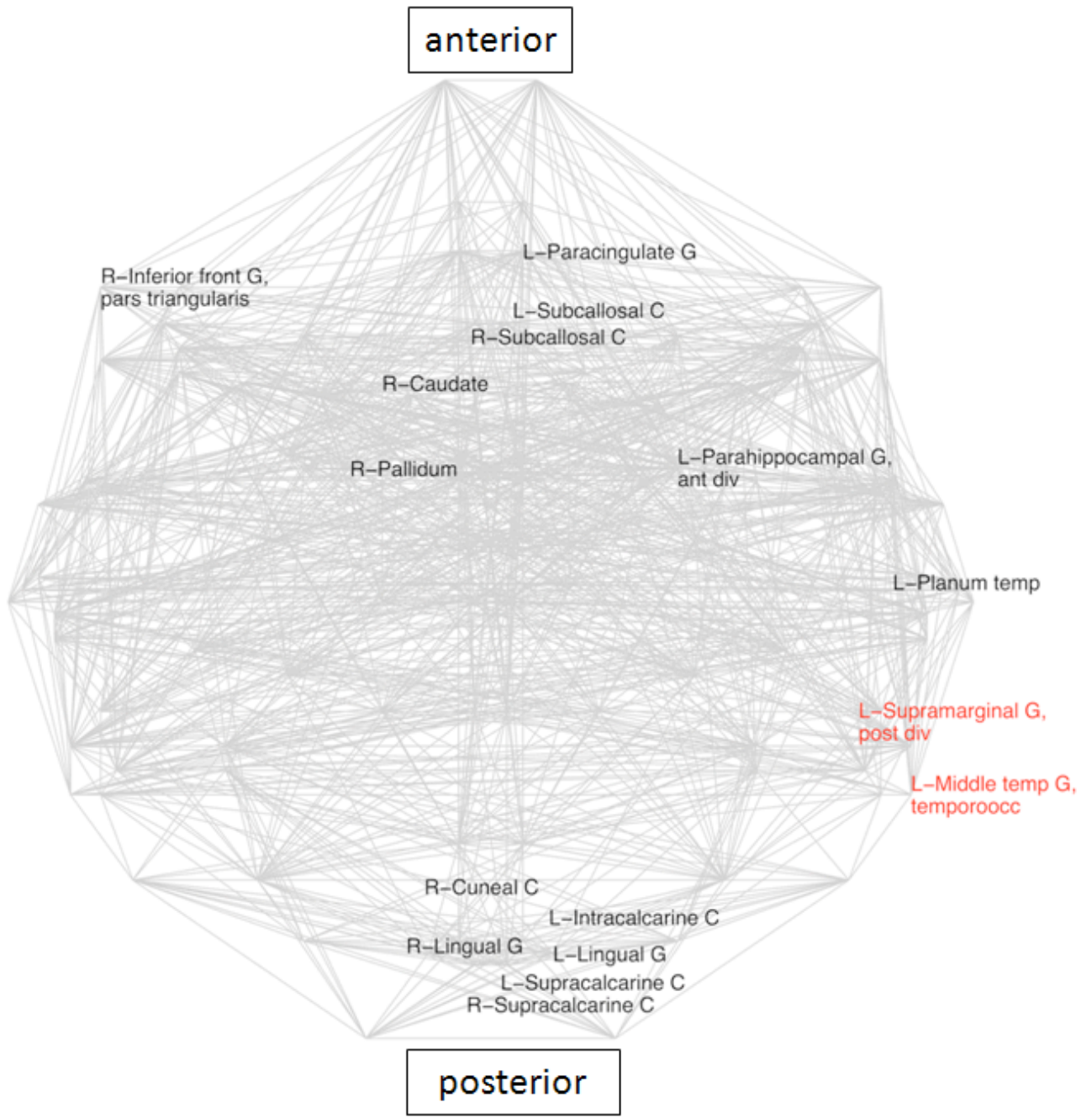

### 3.3. Local circuits

The caudate has been identified in earlier studies (Genovese et al., 2002) as a crucial area involved in LLD, due to its known role in emotion regulation. Analysis of connectivity of the right caudate in the present study between the groups demonstrated the existence of 16 nodes which were specific for the LLD subjects (Figure 3), importantly including the posterior cingulate cortex (PCC) and the Precuneus (PC), which are elements of the DMN. While looking at frequencies of neighbours of the right caudate nucleus (rCN) present in the two groups, six areas occurred more frequently in controls and three were more prevalent in LLD ($z$-score>2 of pooled frequencies were considered as significantly different, see Figure 4). Despite variability in frequencies of hub's occurrences between groups, connections with medial-parietal areas observed in the rCN tended to occur more frequently in LLD patients ( Figure 3).

   -- *Fig 3 about here* –

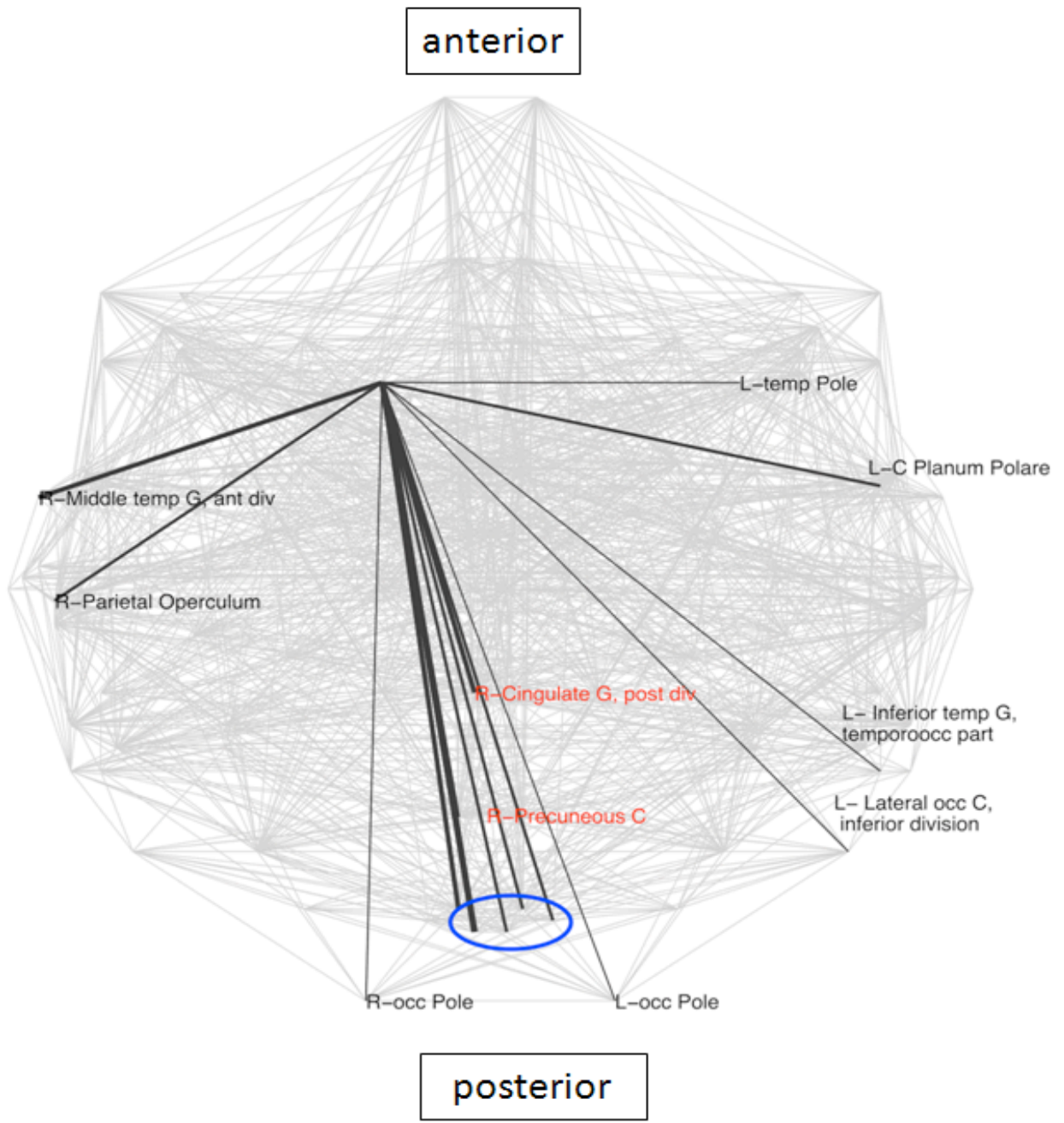

*-- Fig 4 about here –*

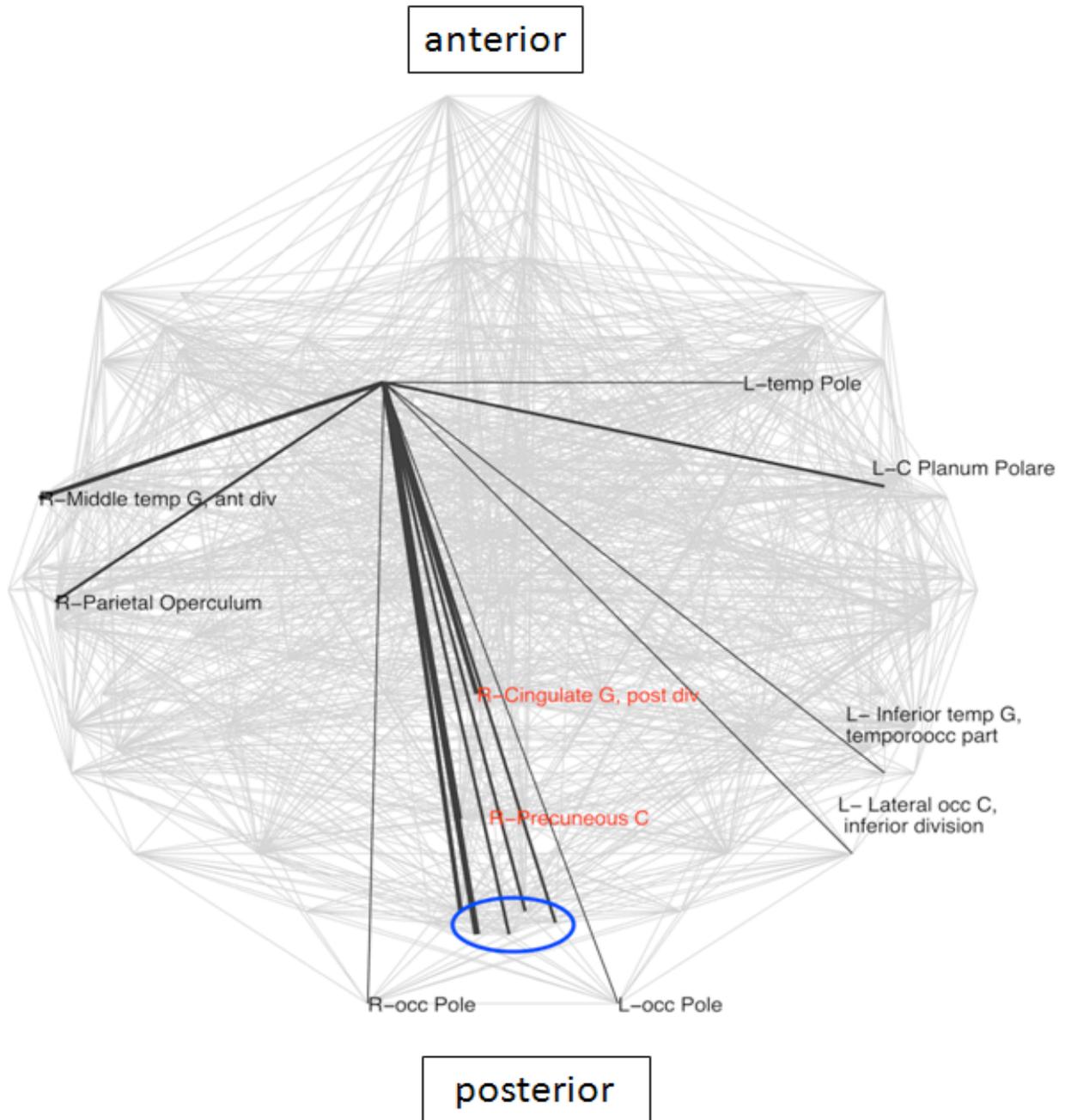

### 3.4. Effect of anti-depressant medications

A previous study (Anand et al., 2005b) reported an up-regulatory effect of selective serotonin reuptake inhibitors (SSRI) such as sertraline on connectivity between the anterior cingulate and limbic regions. To verify that our findings are not due to SSRI activity, we compared the global strength of connectivity (based on CM) and node-related average Euclidean distances for SSRI-takers (see Table 2) and patients not taking these drugs. We found no significant differences for both sub-groups. Based on these analyses, differences observed for controls versus LLD group are unlikely to result from SSRI intake, although due to the small SSRI subgroup we cannot dismiss this effect entirely. Ideally, all depressed subjects would be medication free but the associated ethical concerns with this would be great..

## 4. Discussion

In this study, we showed distinct differences in functional connectivity between LLD subjects and similarly aged healthy controls. First, at the global level, the average correlation strength is higher in LLD. Secondly, spatial properties of individual nodes were altered in LLD: 16 nodes showed a significant difference for average spatial (Euclidean) distance between connected nodes with 14 increased and two reduced distances (Figure 2). Third: the core hubs for LLD comprised the medial PCC and ASG, whereas in controls the MFG was more common. Below we discuss these three major points

### 4.1. Depression: increased or decreased connectivity?

Our analysis of strength of global association between nodes (of raw CM) yielded higher values for LLD. This is in agreement with a number of studies that have reported increased connectivity in depression. However, there have also been papers showing decreased connectivity in this condition. One possible explanation for this discrepancy might be significant methodological differences between studies. For example, some studies use model-free approaches (Greicius et al., 2007;Veer et al., 2010) whereas others use model-based approaches (Bluhm et al., 2009;Sheline et al., 2010;Zhou et al., 2010). Functional connectivity can either be determined in the resting-state (Greicius et al., 2007;Bluhm et al., 2009;Sheline et al., 2010) or while performing tasks (Aizenstein et al., 2009;Grimm et al., 2009;Sheline et al., 2009). For studies with depression patients over 30 years of age, increased connectivity has generally been reported (Greicius et al., 2007;Bluhm et al., 2009) with few studies reporting decreased connectivity e.g. (Veer et al., 2010).

There are only few publications investigating FC in LLD (see e.g. Yuan et al., 2008;Aizenstein et al., 2009). A study reported decreased FC ,(Aizenstein et al., 2009) whereas the study by Zhang et al. in subjects with a wide range of age (Zhang et al., 2011) reported both increased (putamen, frontal and parietal cortex) and decreased (frontal, temporal and parietal cortices) FC. A recent study showed an increased global network integrity metrics based on graph theory (increased efficiency; decreased characteristic path length) and locally for a range of nodes (increased nodal centrality) in freshly diagnosed drug-naive patients (Zhang et al., 2011). The findings from the current study are partially consistent with reports of increased connectivity, at least as revealed at the level of raw cross-correlation matrices.

At nodal level tendencies towards increased connectivity was observed for all types of networks analysed, but these differences did not survive FDR correction. Noteworthy one of the areas with a higher degree (number of neighbours) for binary graphs in LLD was the right anterior cingulate (31.93 ±6.76 vs. 25.44±7.55, t-test p=0.018, uncorrected). In a previous study an increase of connectivity was reported in subgenual cortex, which is a small part of anterior cingulate (Greicius et al., 2007). Therefore it is tempting to hypothesize that an observed tendency towards increase in the number of connections for the anterior cingulate cortex was driven by the subgenual cortex. Psychosurgical interventions specifically target major projections and elements of the DMN such as anterior cingulate cortical tracts connecting it to other structures. In recent years, more refined methods include deep brain stimulation for treatment-resistant forms of depression. Interestingly, the subgenual cingulate cortex, one of the regions targeted by this technique for depression symptoms relief (Mayberg et al., 2005) was also found to be characterized by an increased FC in depression patients (Greicius et al., 2007).

**4.2. Late life depression: increase of correlation length**

This is the first study to report increased average Euclidean distance (ED) between many nodes in LLD (Fig 2). The increases in geometrical distances of average connections in LLD suggest the prevalence of long connections implying more intense communications between large and widely distributed components of the brain networks such as the DMN. Indeed, a recent study (Zhang et al., 2011) suggested that diminished average $L$ in major depression is linked to an increased number of long-range connections. In addition, elements of the DMN were characterized by higher centrality metrics. An overall increase in long-distance connections (observed in this study) could be explained by an up-regulation of DMN activity, as areas displaying higher ED values were core components of the posterior part of the DMN. Amongst the areas with up-regulated mean ED is the right caudate. This region also showed a more diffuse pattern of connectivity in a previous seed-based analysis of the same data (Sheline et al., 2009). The observed rise in ED may be regarded as another altered feature of connectivity related to the caudate associated with LLD.

**4.3. LLD and core hubs**

The results of this study suggest that LLD spares general organization of FC, at least in relation to the aggregate topological measures used. This is in contrast to neurodegenerative diseases such as Alzheimer's disease that show higher characteristic path lengths in FC and decrease in small-worldness properties (de Haan et al., 2009). In general, many neurodegenerative disorders seem to target specific elements of the brain that are considered to be critical parts of its topology (Buckner et al., 2009). We therefore specifically investigated the connectivity pattern and structure of the 15 top hubs (Table 3). Changes in composition of the core of hubs were observed, with LLD individuals having a higher frequency of medial PCC and one parietal structure anterior supramarginal gyrus (ASG), whereas controls had a higher frequency in the MFG. Within the set of core-hubs, a similar pattern was also determined by Kenny and co-workers (Kenny et al., 2010). PCC is a crucial component of the DMN and is thought to play a role in interpreting other people's feelings and envisaging the future (Buckner et al., 2008). Importantly, it is part of the limbic system and disturbances in its connectivity, especially with the frontal cortex, were related to psychiatric diseases including depression and schizophrenia (Buckner et al., 2008;Johnson et al., 2009). In line with connectivity abnormalities, a lower inhibition of DMN activity was shown in attention-demanding tasks in relation to depression (Anand et al., 2005a;Greicius et al., 2007;Auer, 2008;Buckner et al., 2008). Importantly a recent paper more specifically pointed to the significance of over-activity of the posterior medio-parietal complex comprising the PCC in major depression (Sheline et al., 2009). The role of supramarginal gyrus (SG) in LLD is more difficult to interpret, however this structure lies in close proximity to parietal components of the DMN. The study by Buckner and co-workers (Johnson et al., 2009) identified SG as one of the critical cortical hubs, similarly in a DTI (Buckner et al., 2009) and morphometric connectivity study (Gong et al., 2009). Importantly, the authors also noted an overlap of the network comprising SG with a network containing PCC/PC (core constituents of the DMN).

**4.4. Experimental group composition and limitations of the study**

The first possible concern about this study is the definition and composition of the patient group. Although this group was characterized by a spread in clinical characteristics (e.g. age of onset and number of depression episodes) the patients shared the features which were in the centre of our attention: the occurrence of depression in later life, rather than late-onset depression. Despite the variance in age of onset, all patients had suffered an initial episode of depression followed by remission with then at least another one episode in later life. Another point is that they were not depressed at the time the study was performed enabling us to look at patients state rather than trait. Therefore these findings may reflect features which are either a consequence of the previous disease or are constituent part of the brain organization in subjects vulnerable to depression. There is another possible concern. Although subjects were not currently depressed we did not include a specific measure of severity of anxiety symptoms and it is therefore possible that some of the changes in connectivity we identified reflected comorbid anxiety symptoms. Last but not least: there were medications taken by a part of patient group. Ideally, all depressed subjects would be medication-free: however for a study looking for a long term effects of a disease, it is very difficult to recruit a sufficiently large group of patients completely free of medications. In addition we addressed a possible effect of SSRIs and found no significant impact on our findings.

It should be stressed that there are several potential confounds to the resting state signal such as, for example, physiological noise (both respiratory and cardiac related). Over the last decade various approaches to remove potential noise have been assessed, but this still remains a key area of investigation (for reviews: Birn, 2012;Snyder and Raichle, 2012). In this study, we carried out high-pass temporal filtering and global signal removal to account for potential noise in our data. More recently, other studies have regressed white matter and CSF signal and included motion parameters in their analysis. These methods are receiving growing attention and are being used (see e.g. Liao et al., 2010;Zuo et al., 2012) in addition to the methods that we used (e.g. Lynall et al., 2010;Sanz-Arigita et al., 2010).

In addition recent studies have applied corrections to the extracted BOLD signal to account for potential effects caused by brain atrophy (see e.g. Binnewijzend et al., 2012;Voets et al., 2012). Although atrophy is a potential confound, we did not observe significant differences in the brain volume between the groups, therefore we conclude that levels of atrophy might not be a factor that could explain FC differences between LLD patients and controls.

We believe that despite these limitations the study gives a valuable insight into the characteristics of the state of the brain affected by a relatively long history of a mental disease. It provides new information and/or corroborating previous findings or suggestions.

### 4.5. Conclusions

This is the first functional connectivity study showing that in LLD specific brain areas are characterised by higher correlation lengths (Euclidean distances between nodes with correlated activity). In line with the above notion, the average functional correlation strength is higher in LLD. In contrast, clustering coefficient, characteristic path length, and modularity in tresholded binary networks were unaffected in LLD. In LLD, connectivity with the caudate nucleus (right) showed a more diffuse pattern and linked closer to the core elements of the DMN. In conclusion, this study reports some interesting findings of altered connectivity in LLD and highlights the potential use of resting-state functional connectivity in characterising LLD.

**Conflict of interest statement**

The authors declare that the research was conducted in the absence of any commercial or financial relationships that could be construed as a potential conflict of interest.

**Acknowledgments**
The authors would like to thank Jose Marcelino and Dr Saad Jbabdi for writing scripts in Matlab and bash shell, valuable advice, and help with FSL. We are also grateful to Dr Cheol Han for the help in using his Matlab script to implement permutation-FDR method for multiple comparison correction. Marcus Kaiser acknowledges support by the WCU program through the National Research Foundation of Korea funded by the Ministry of Education, Science and Technology (R32-10142), by EPSRC (EP/G03950X/1), and by the CARMEN e-science project (http://www.carmen.org.uk ) funded by EPSRC (EP/E002331/1).  John O'Brien declares the support from UK NIHR Biomedical Research Centre for Ageing and Age-related disease award to the Newcastle upon Tyne Hospitals NHS Foundation Trust. Eva Kenny was supported by a Medical Research Council UK capacity building studentship.

**Figure legends:**

**Figure 1.** Major steps of functional connectivity analysis. Parcellation of the brain into areas based on the anatomical atlas and extraction of demeaned time series BOLD signal from each area (A), construction of correlation matrices (B) thresholding and binarization of correlation matrices; generation of binary adjacency matrices (C) visualized in (D), analysis of topology and microcircuit patterns (E). In the blue boxes are the names of main software tools used at relevant stages. See Methods for further details.

**Figure 2.** Areas with significantly different average Euclidean distances to its neighbours (inter-group differences), superimposed on the whole brain connectivity projected onto one axial plane, averaged for all subjects in each group (pale grey lines), FDR: 5% -corrected. LLD-related increases in black, decreases in red. Abbreviations: R\L: right\left hemisphere, front: frontal, G: gyrus, inf: inferior, occ: occipital, temp: temporal.

**Figure 3.** Connections specific for the right caudate in LLD group. These connections are superimposed on the whole brain connectivity (projected onto one axial plane), averaged for all subjects in each group (pale grey lines). The thickness of black lines is proportional to the frequency of occurrence of a particular caudate connection in relation to the total number of connections in each group. Depicted in red are core elements of the default-mode network (DMN). Abbreviations: R\L: right\left hemisphere, front: frontal, G: gyrus, inf: inferior, occ: occipital; blue oval depicts a cluster of closely located structures of the primary visual cortex, consisting of bilateral cuneal and supracalcarine cortices as well as left lingual and intracalcarine cortices

**Figure 4.** Areas with significantly different frequencies of right caudate connections between the groups (z-score > 2) superimposed on the whole brain connectivity (projected onto one axial plane), averaged for all subjects in each group (pale grey lines). LLD-related increases in black, decreases in red (note: only connections shared in the two groups were taken into account), Abbreviations: R\L: right\left hemisphere, G: gyrus, front: frontal, occ: occipital.